\documentclass[%
 aip,
 jmp,%
 amsmath,amssymb,
reprint,%
]{revtex4-1}

\usepackage{graphicx}
\usepackage{dcolumn}
\usepackage{bm}
\usepackage{hyphenat}

\begin{document}

\preprint{AIP/123-QED}

\title[]{Straintronic magneto-tunneling-junction based ternary content addressable memory}

\author{S. Dey Manasi}
\affiliation{%
Department of Electrical and Computer Engineering, University of Illinois at Chicago, Chicago, IL 60607  USA
}

\author{M. M. Al Rashid}%
\affiliation{
Department of Mechanical and Nuclear Engineering, Virginia Commonwealth University USA, Richmond, VA 23284, USA
}%

\author{J. Atulasimha}
\affiliation{%
Department of Mechanical and Nuclear Engineering, Virginia Commonwealth University USA, Richmond, VA 23284, USA
}
%
\author{S. Bandyopadhyay}
\affiliation{%
Department of Electrical and Computer Engineereing, Virginia Commonwealth University, Richmond, VA 23284, USA
}

\author{A. R. Trivedi}
\affiliation{%
Department of Electrical and Computer Engineering, University of Illinois at Chicago, Chicago, IL 60607, USA
}

\date{\today}

\begin{abstract}
Straintronic magneto-tunneling junction (s-MTJ) switches, whose resistances are controlled with voltage-generated 
strain in the magnetostrictive free layer of the MTJ, are extremely energy-efficient switches that would dissipate a 
few aJ of energy during switching. Unfortunately, they are also relatively
error-prone and have low resistance on/off ratio. This suggests that as computing elements,
they are best suited for non-Boolean architectures. Here, we propose and analyze a
ternary content addressable memory implemented 
with s-MTJs and some transistors. 
It overcomes challenges encountered by traditional all-transistor implementations,
resulting in exceptionally high cell density and an energy-delay product that is orders of magnitude lower. 
\end{abstract}
\maketitle

\section{INTRODUCTION}
The primary threat to continued downscaling of electronic devices envisaged in Moore's law \cite{moore}
is the excessive energy dissipation that takes place in the device during switching.
Straintronic magneto-tunneling junctions (s-MTJ)  are among the most energy-efficient three-terminal resistance switches 
extant 
\cite{kuntal1, kuntal2, salehi1, salehi2}. Unfortunately, they are also relatively error-prone 
\cite{atul_sci_rep,kam,salehi,mamun,mamun4} and have low resistance 
on/off ratios. The switching 
error probability is typically larger than 10$^{-9}$ at room temperature \cite{mamun,mamun4,ayan_sci_rep} which makes it problematic to
utilize them in Boolean logic. This has turned attention to non-Boolean computing paradigms \cite{andras1, andras2}, which may be more 
forgiving of errors and do not always demand high resistance on/off ratios. Here, we explore one such application, 
namely Ternary Content-Addressable Memory (TCAM) and show that replacing transistors with s-MTJ results in significant energy saving and 
increased 
cell density. The low on/off ratio does not inhibit circuit operation, although a higher on/off ratio would be desirable.

TCAM is useful for high-speed parallel data processing. It finds application in platforms such as packet forwarding in network routers 
\cite{amit1}, image encoding \cite{amit2}, parametric curve extraction \cite{amit3}, and Hough transformation \cite{amit4}. It
compares input search data against a table of stored data to return the memory address of fully or partially matching data. Each TCAM cell has 
three states in its search and storage bit: `0', `1' and `X' (don't care). The ``don't care'' state allows masking, i.e., a match regardless of the 
storage and/or search data bit. Key challenges in a large scale TCAM are to achieve a high cell density and low standby power dissipation. 
Conventional CMOS-based TCAM cells consume large areas on a chip. Although CMOS scaling improves the cell density, the standby 
power dissipation deteriorates \cite{amit5}. On the other hand, an s-MTJ based 
TCAM can overcome these challenges and achieve a very high cell density along with little or no standby power dissipation.

\section{SKEWED STRAINTRONIC MAGNETO-TUNNELING JUNCTION (s-MTJ)}

An s-MTJ is a standard MTJ (fixed layer-spacer-free layer) with one difference. The free layer is a magnetostrictive nanomagnet in
elastic contact with an underlying poled piezoelectric thin film of thickness $a$ as shown in Fig. 1(a). Square electrodes of edge $L$
($\approx a$), separated by a distance 
$d$ ($L \leq d \leq 2L$), are delineated on the piezoelectric surrounding the MTJ stack. The bottom of the conducting substrate is grounded. 
The electrode
`1' is used to read the s-MTJ resistance by passing a current to ground. Application of a voltage across the piezoelectric film
using the electrode pair `2' shown in Fig. 1(a) generates biaxial strain in the film (compresssion along the line joining the electrode pair 
and tension perpendicular to it, or vice versa, depending on the polarity of the voltage),
which is partially or fully transferred to the soft layer of the s-MTJ in elastic contact with the film. This rotates its
 magnetization via the Villari effect 
\cite{cui1, cui2, cui3} and changes the s-MTJ resistance, realizing the action of a switch. 
A tiny amount of voltage 
$V$ (few mV) is required to rotate the magnetization through a large angle and change the s-MTJ resistance substantially
if the piezoelectric film is $\sim$100 nm thick, 
resulting in a switching energy dissipation 
$CV^2$ ($C$ = capacitance associated with charging the piezoelectric, which is 1-2 fF) of a few tens of aJ \cite{hasnain,noel}. The internal
energy dissipation within the magnetostrictive free layer due to Gilbert damping is negligible \cite{kuntal1}. 

\begin{figure}[!ht]%
 \centering
  \includegraphics[width=3.6in]{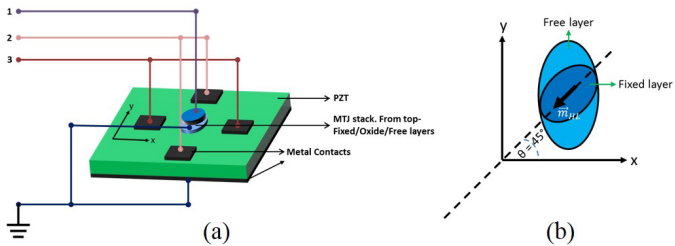}
 \caption{(a) A 4-terminal s-MTJ switch showing the MTJ stack, the piezoelectric layer and the electrodes.
 (b) The top view of the free and fixed layers of the MTJ. The major axes of the two ellipses subtend an
 angle of 45$^{\circ}$ between themselves.}
  \label{fig:comparator}
\end{figure}

The s-MTJ operation 
has been experimentally demonstrated \cite{li,wang}. Here, we first show
that an s-MTJ can be engineered to produce very unusual device characteristics and then show that such device characteristics 
 elicit TCAM behavior.

Consider a ``skewed'' s-MTJ where
the major
axes of the fixed and free layers  subtend an angle of 45$^{\circ}$ between them as shown in Fig. 1(b). The fixed layer 
is implemented with
a synthetic anti-ferromagnet (SAF) to reduce the dipole interaction with the free layer, {\it but not completely eliminate 
it}. Because of shape anisotropy, the magnetization orientations of both layers will lie along the respective major axes of the ellipses,
but owing to the remanent dipole interaction, the angle between them will be obtuse rather than acute (see Fig. 1(b)). When the 
free layer is strained by the voltage applied at the electrode pairs `2', its magnetization begins to rotate. The remanent dipole interaction, 
will make it rotate {\it clockwise} in Fig. 1(b) so as to always {\it
increase} the angular separation between the magnetizations of the two layers. We have simulated the magnetization rotation under strain 
in the presence of 
thermal noise 
using stochastic Landau-Lifshitz-Gilbert (s-LLG) simulations (see later description). Out of 10$^6$ switching trajectories 
simulated, not a single one rotated anti-clockwise, showing that the clockwise rotation is overwhelmingly preferred.

Strain can usually rotate magnetization by up to 90$^{\circ}$ (from the major to the minor axis of the ellipse)
although larger rotations are possible 
under special circumstances \cite{kuntal_sci_rep, ayan_apl}. Initially, before the application of strain, 
the magnetizations of the fixed and free layers subtend an angle $\theta$ = 135$^{\circ}$ as shown in Fig. 1(b). Upon application
of strain, the magnetization begins to rotate clockwise and 
$\theta$ gradually increases from 135$^{\circ}$ to 225$^{\circ}$. 
The MTJ resistance depends on $\theta$ according to \cite{camsari}
\begin{equation}
{{R(\theta) - R_P}\over{R_{AP} - R_P}} = {{1 - cos \theta}\over{\chi (1 + cos \theta) + 2}} ,
\label{resistance}
\end{equation}
where $R_{P(AP)}$ is the MTJ resistance when the magnetizations of the fixed and free layers are parallel (anti-parallel), $R(\theta)$ is 
the resistance when the angular separation between the magnetizations is $\theta$, and $\chi = \left ( R_{AP} - R_P \right )/R_P$. Since $\theta$ varies
between 135$^{\circ}$ and 225$^{\circ}$, the conductance of the 
MTJ (or current flowing through terminal `1' at a fixed bias) plotted as a function of the voltage applied at terminal `2' (which generates 
the rotation) will exhibit a ``valley''. The bottom 
of the valley corresponds to $\theta = 180^{\circ}$ when the MTJ resistance becomes maximum. 

We can alter the stress distribution in the free layer of the s-MTJ by applying an additional voltage across the piezoelectric with
 a third pair of electrodes `3' shown in Fig. 1(a). This will allow us to shift the position of the valley bottom in the transconductance 
 characteristic $I_1$ versus $V_2$ ($I_n$ is the current through the 
$n$-th terminal at a fixed bias and $V_n$ is the voltage applied at the $n$-th terminal). Thus, we have a 4-terminal switch with terminals
`1', `2', `3' and ground, where the current between `1' and ground is changed with a voltage applied to `2' and the transfer characteristic 
associated with this change can be modulated with a voltage applied at terminal `3'. 

When both electrode pairs `2' and `3' are activated, the strain
 distribution in the piezoelectric (and hence in the free layer of the s-MTJ)
 becomes complex. Exact strain profiles can be calculated with three dimensional finite element analysis (e.g. with COMSOL Multiphysics package)
 as in
 \cite{cui1,wang}, but in order to keep the analysis tractable, we will assume that activating an electrode pair generates only
 uniaxial stress along the line joining that pair. Note that if anything, this over-estimates the stress required to produce 
 a given rotation $\theta$, and is hence conservative. The sign of the uniaxial stress (tensile or compressive) depends on the polarity of the 
 voltage. If we activate electrode pair `2', then we will generate uniaxial stress along the major axis of the elliptical
 free layer of the s-MTJ (compressive or tensile 
 depending on the voltage polarity at `2'), whereas if we activate electrode pair `3' we will generate uniaxial stress along the 
 minor axis of the free layer. We have assumed that the free layer is made of Terfenol-D which has a positive and large magnetostriction 
 coefficient 
 (900 ppm). Compressive stress along any direction in the free layer will rotate its magnetization away from that direction
(maximum rotation is 90$^{\circ}$) 
 while tensile stress 
 will keep it aligned along that direction. This allows us to control the angle $\theta$ with voltages at `2' and `3'.
 
 We have computed $\theta$ versus the voltage $V_2$ (assuming $V_3$ = 0) at 0 K temperature (no thermal noise) using the 
 Landau-Lifshitz-Gilbert 
 equation which yields the magnetization orientation of the free layer as a function of time $t$ under the influence of voltage generated 
 stress:
 \begin{equation}
 {{d {\vec M} (t)}\over{dt}} = - \gamma {\vec M}(t) \times {\vec H}_{eff} (t) - {{\alpha \gamma}\over{M_s}} \left [ \vec M (t) \times 
 \left ( \vec M (t) \times {\vec H}_{eff} (t) \right ) \right ] ,
 \label{llg}
 \end{equation}
 where $M_s$ is the saturation magnetization of the free layer material, $\gamma$ is the gyromagnetic ratio, $\alpha$ is the Gilbert damping constant 
 in the free layer, and ${\vec H}_{eff} (t)$ is the effective magnetic field experienced by the free layer at any time $t$ and is given by
 \begin{equation}
 {\vec H}_{eff} (t) = {\vec H}_{dipole}  + {\vec H}_{shape}(t) + {\vec H}_{stress}(t) + {\vec H}_{thermal}(t),
 \label{eff-field}
 \end{equation}
 where ${\vec H}_{dipole}$ is the (constant) dipole field exerted by the fixed layer, ${\vec H}_{shape}(t)$ is the field due to shape anisotropy, 
 ${\vec H}_{stress}(t)$ is the field generated by stress, and ${\vec H}_{thermal}(t)$ is the random field due to 
 thermal noise. Expressions for these fields are given in ref. \cite{salehi1, kuntal2}. Stress is generated in the 
piezoelectric substrate by activating a shorted electrode pair with a voltage $\cal V$. The resulting stress is assumed to 
be uniaxial along the line joining the centers of the electrodes in the activated pair.
The voltage $\cal V$ generates a vertical electric field of  ${\cal V}/a$ in the piezoelectric substrate. Following Cui, et al. \cite{cui1},
we will assume that a vertical electric field of 37 kV/m is required to produce a uniaxial stress of 1 MPa in the substrate along the line 
joining the electrodes in the activated pair. This stress is assumed to be fully transferred 
to the soft magnetic layer of the MTJ resting on top of the substrate.
 A negative voltage generates tensile stress and a positive voltage compressive stress because of the direction in which the piezoelectric 
 film has been poled.
 Equation (\ref{llg}) is solved for various $\cal V$-s until steady state is reached and that yields the orientation of the free 
 layer's magnetization
 as a function of the $\cal V$-s and hence $\theta$ versus $V_2$ for a fixed $V_3$. This result is plotted in Fig. 2 (a)
 for 0 K and 300 K tempatures, assuming $V_3$ = 0 V and ${\vec H}_{dipole} = 7.05 mT$ directed along the major axis of the fixed layer. 
 The dispersion in the 300 K result is due to thermal noise.
 The parameters assumed for the free layer (material Terfenol-D)
 are given in Table I. For the MTJ, we assumed the spacer layer to be made of MgO of thickness 1 nm. For this thickness, the resistance-area
 product of the MTJ is about 10 $\Omega$-$\mu m^2$ \cite{isogami}. If the thickness is increased to 2 nm, the resistance-area product 
 increases to 8000 $\Omega$-$\mu m^2$.
 
 \begin{table}
 \caption{Parameters for the free layer}
 \begin{tabular}{|c|c|}
 \hline
 Saturation magnetization ($M_s$) & 8$\times 10^5$ A/m \\
 Major axis dimension & 80 nm \\
 Minor axis dimension & 60 nm \\
 Thickness & 15 nm \\
 Magnetostriction coefficient & 900 ppm \\
 Gilbert damping constant & 0.1 \\
 \hline
 \end{tabular}
 \end{table}
 
 \begin{figure}[!ht]%
 \centering
  \includegraphics[width=3.6in]{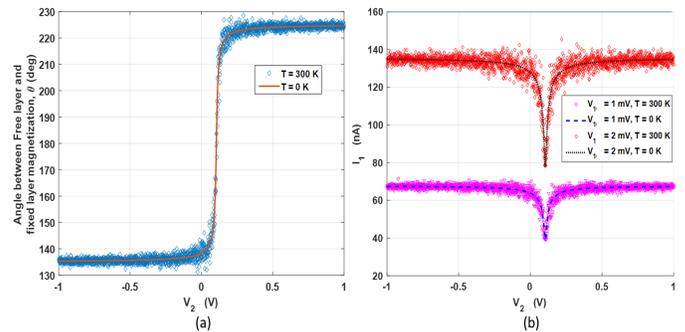}
 \caption{(a) The angle $\theta$ between the magnetizations of the free and fixed layers plotted as a function of the voltage $V_2$ 
 applied at the 
 electrode pair `2'. The voltage $V_3$ = 0 V and the dipole field ${\vec H}_{dipole}$ experienced by the free layer
 is assumed to be 7.05 mT directed along the major axis of the fixed layer.  The results are plotted for two different 
 temperatures.
 The dispersion in the 300 K curve is due to thermal noise. (b) The transfer characteristic $I_1$ versus $V_2$ for two different temperatures 0 K and 300 K. The results are plotted for 
 $V_3$ = 0 and ${\vec H}_{dipole}$ = 7.05 mT directed along the major axis of the fixed layer.}
  \label{}
\end{figure}
 
 We then use Equation (\ref{resistance}) to extract the s-MTJ resistance $R_{\theta}$ versus $V_2$ from the $\theta$ versus $V_2$ 
 realtion in Fig. 2(a) and plot 
 the transfer characteristic $I_1 (= V_1/R_{\theta})$ versus $V_2$ (at
 0 K and 300 K temperatures)
 in Fig. 2(b) for two different values of $V_1$. Note that this characteristic has a notch or valley. Note also that there is no significant 
 difference between the 0 K and (average of) 300 K results. Therefore, in the rest of this paper, we will present the 0 K results, 
 {\it noting that the 
 300 K results will not be significantly different}.
 In Fig. 3, we show how the transfer characteristics depend on the dipole field strength, assuming that the temperature is 0 K.
 
 \begin{figure}[!ht]%
 \centering
  \includegraphics[width=3.6in]{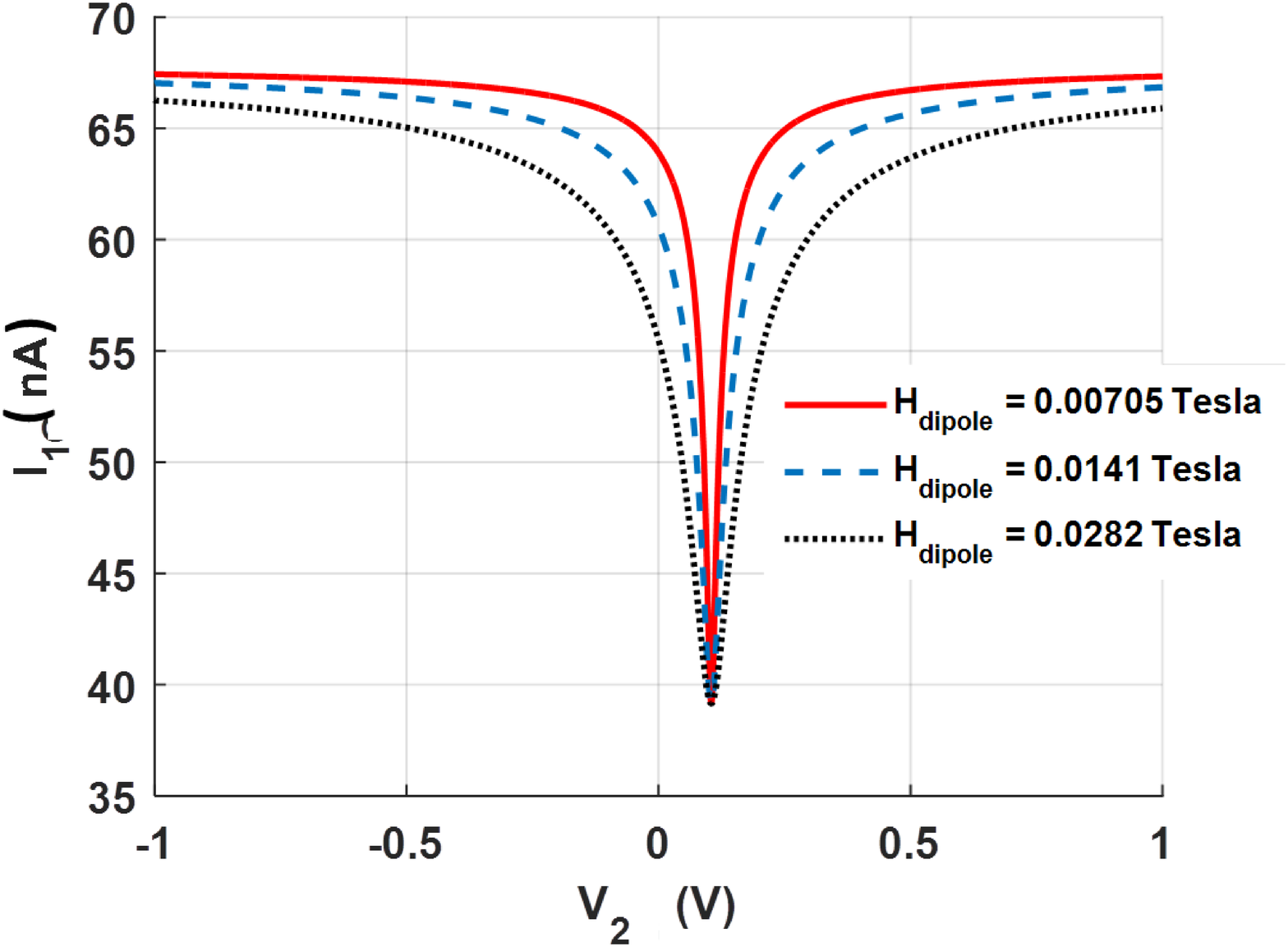}
 \caption{The transfer characteristic plotted at 0 K temperature for three different values of the 
 dipole field ${\vec H}_{dipole}$ directed along the amjor axis of the fixed layer, assuming $V_3$ = 0 V.}
  \label{}
\end{figure}

\begin{figure}[!ht]%
 \centering
  \includegraphics[width=3.6in]{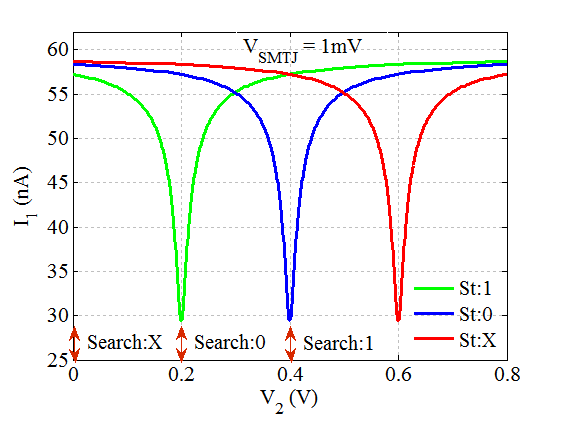}
 \caption{The current $I_1$ through the cell s-MTJ at varying search bit potentials $V_2$ .
 The dipole field ${\vec H}_{dipole}$ is assumed to be 7.05 mT directed along the major axis of the fixed layer.}
  \label{}
\end{figure}

 In Fig. 4, we plot the transfer characteristic $I_1$ versus $V_2$ at 0 K temperature for three different values of $V_3$. Clearly, the position 
 of the notch can be shifted 
 around with the voltage $V_3$ which generates an additional uniaxial
 stress (negative voltage tensile and positive voltage compressive) along the line joining the electrode pads `3'. This makes it a 4-terminal
 switch.

\section{s-MTJ-based Dynamic TERNARY CONTENT ADDRESSABLE MEMORY (TCAM)}
In a skewed s-MTJ, the current $I_1$ between the free and fixed layers can be controlled by the gate voltages at $V_2$ and $V_3$ [Fig. 4]. 
At any given value of $V_1$, $I_1$ is lowest when  $V_2$ and $V_3$ `match', meaning that they 
obey the relation $V_3 = V_2 + V_F$, where $V_F$ is a fixed voltage that we call the `offset voltage'. 
The current $I_1$ increases steeply when  $V_2$ and $V_3$ deviate from the `match' condition. Therefore, the current through skewed s-MTJ 
characterizes similarity between the gate voltages $V_2$ and $V_3$. Moreover, a current-based similarity index in skewed s-MTJ is suitable 
for an easier inter-cell aggregation and for evaluating similarity between large scale vectors/patterns. When multiple skewed s-MTJs are 
arranged in parallel, the column current aggregates the similarity index (i.e., the s-MTJ current $I_1$) from each cell. Therefore, the column current 
evaluates similarity between two vectors/patterns, each applied at the $V_2$ and $V_3$ nodes of the column, respectively. We have exploited 
this associative processing capability of s-MTJ in TCAM design.

\begin{figure}[!ht]%
 \centering
  \includegraphics[width=3.6in]{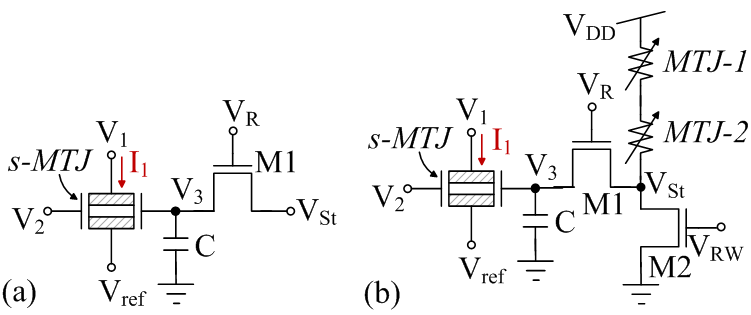}
 \caption{(a) s-MTJ based dynamic TCAM Cell. (b) s-MTJ-based dynamic TCAM cell with local refresh using MTJs.}
  \label{}
\end{figure}

\begin{figure}[!ht]%
 \centering
  \includegraphics[width=2.1in]{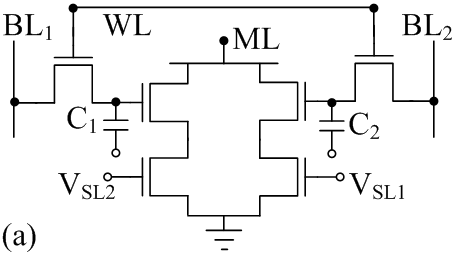}
  \includegraphics[width=3.6in]{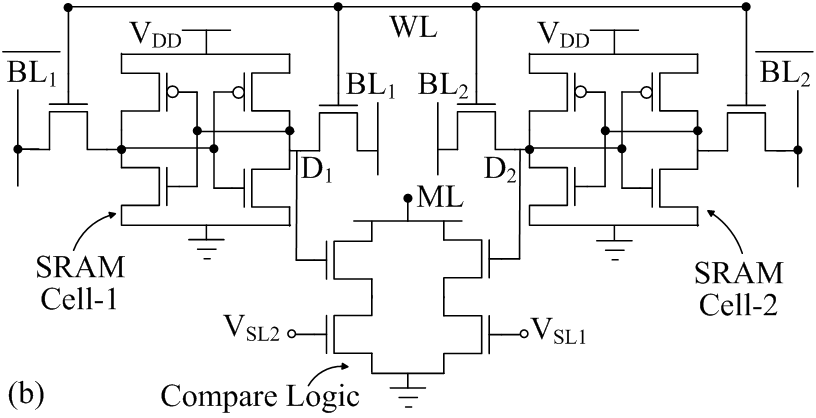}
 \caption{(a) 6T CMOS-based dynamic TCAM. (b) 16T CMOS-based static TCAM.}
  \label{}
\end{figure}

Fig. 4 discusses the encoding scheme for implementing match operation of a TCAM cell through a single skewed s-MTJ. In Fig. 4, s-MTJ current 
($I_1$) is shown at varying search bit potentials (i.e., $V_2$ potential) and at varying stored bits (i.e., $V_3$ potential). The search bits 
`X', `0', and `1' are encoded as 0 V, 0.2 V, and 0.4 V, respectively. The store bits `1', `0', and `X' center the valley peak to 0.2 V, 0.4 V, 
and 0.6 V, respectively. In the encoding scheme, a high s-MTJ current (i.e., a lower resistance in the s-MTJ) indicates a match between the 
stored and search bit. If the stored bit in a cell is `X', current through the s-MTJ is high at all search bits `0', `1' \& `X', thereby 
ignoring (masking) the search bit. Similarly, when the search bit is `X', a high current is induced in the s-MTJ indicating a match irrespective 
of the stored bit. Therefore, the skewed s-MTJ significantly reduces the complexity of match operation in a TCAM. 

\begin{figure}[!ht]%
 \centering
  \includegraphics[width=3in]{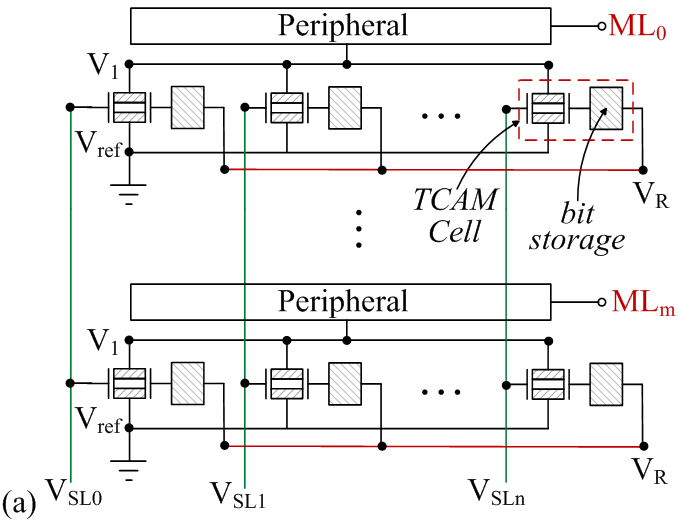}
  \includegraphics[width=2.8in]{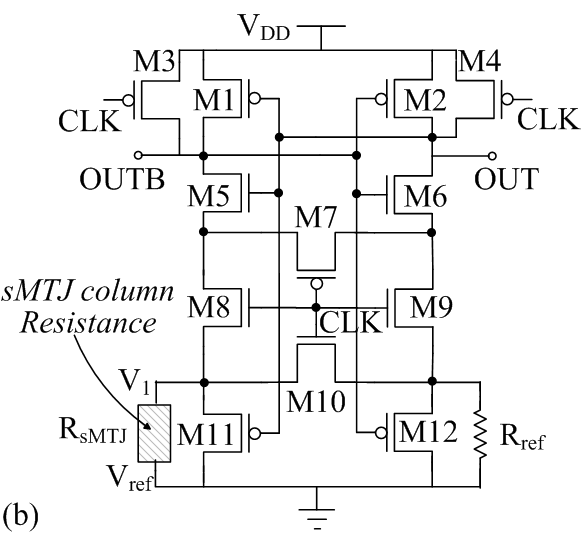}
 \caption{(a) s-MTJ-based TCAM array. (b) Column Sense Amplifier.}
  \label{}
\end{figure} 

The cell schematics for an s-MTJ-based dynamic TCAM is shown in Fig. 5(a). The cells exploit high parasitic capacitance at $V_3$ node for a 
dynamic storage of the storage bit. Note that the capacitance at $V_3$ is high due to an underlying high dielectric constant 
($\epsilon > $ 1000) piezoelectric layer. The parasitic capacitance can be further enhanced by thinning down the piezoelectric layer, 
and/or by increasing the contact area of $V_3$ electrode atop the piezoelectric layer. The parasitic capacitance is charged through the 
access NMOS transistor M1. The cell in Fig. 5(b) supports local refresh of the storage bit, i.e., the storage potential is generated within 
the cell using MTJ-1, MTJ-2, and M2. MTJ-1 and MTJ-2 are standard MTJs (not skewed) and can be switched by spin polarized current 
generating spin transfer torque or domain wall motion.  MTJs locally store the storage bit and refresh the storage 
potential at $V_3$ node when read and access transistors (M1 \& M2) are activated. Both MTJs are programmed at high resistance ($R_H$) 
to store `1' and both at low resistance ($R_L$) to store `X'. One of the MTJs is programmed at $R_H$ and the other at $R_L$ to store `0'. 
The MTJs are designed such that MTJ-2 has a slightly higher critical switching current ($I_{C}$) than MTJ-1. To program both MTJs at 
low resistance, a positive programming voltage magnitude is applied at the $V_{DD}$ node [Fig. 5(b)]. To program both MTJs at high resistance, 
programming voltage polarity is reversed from the previous case (i.e., a negative $V_{DD}$). To write MTJ-1 at $R_{H}$ and MTJ-2 at $R_{L}$, 
first both MTJs are programmed at $R_{L}$. Then a programming pulse of negative polarity whose width is greater than the switching time of MTJ-1 but 
less than the switching time of MTJ-2 is applied at $V_{DD}$ node. Since the critical switching current of MTJ-1 is lower than 
that of MTJ-2, MTJ-1 switches to 
$R_{H}$ while MTJ-2 remains at $R_{L}$. The cell in Fig. 5(a) is designed for a global refresh, i.e., refresh potentials are supplied 
externally to the array (as in a DRAM). Note that a local refresh in Fig. 5(b) does not interfere with the regular search operation, and 
therefore is useful for improving performance and mitigating the complexity of refresh operation. Meanwhile, the global refresh-based cell in 
Fig. 5(a) also reduces cell area. 

\begin{figure}[!ht]%
 \centering
  \includegraphics[width=3.6in]{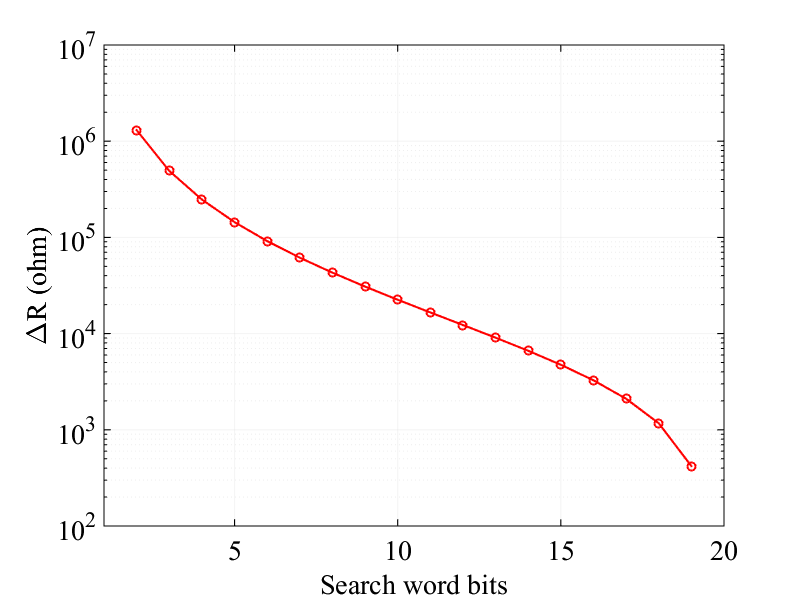}
 \caption{Differential resistance between `worst case match' and `worst case one-bit mismatch' with increasing word length.}
  \label{}
\end{figure}

    \begin{figure}[!ht]%
    \centering
    \includegraphics[width=3.6in]{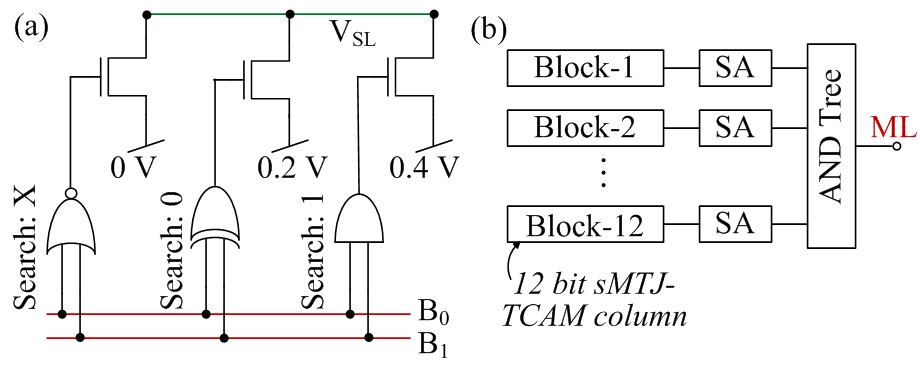}
    \caption{(a) Decoder Circuitry for the search bit. (b) Combination of multiple s-MTJ column blocks to process a 144 bit word.}
    \label{}
    \end{figure} 

        \begin{figure*}[!ht]%
         \centering
          \includegraphics[width=7.2in]{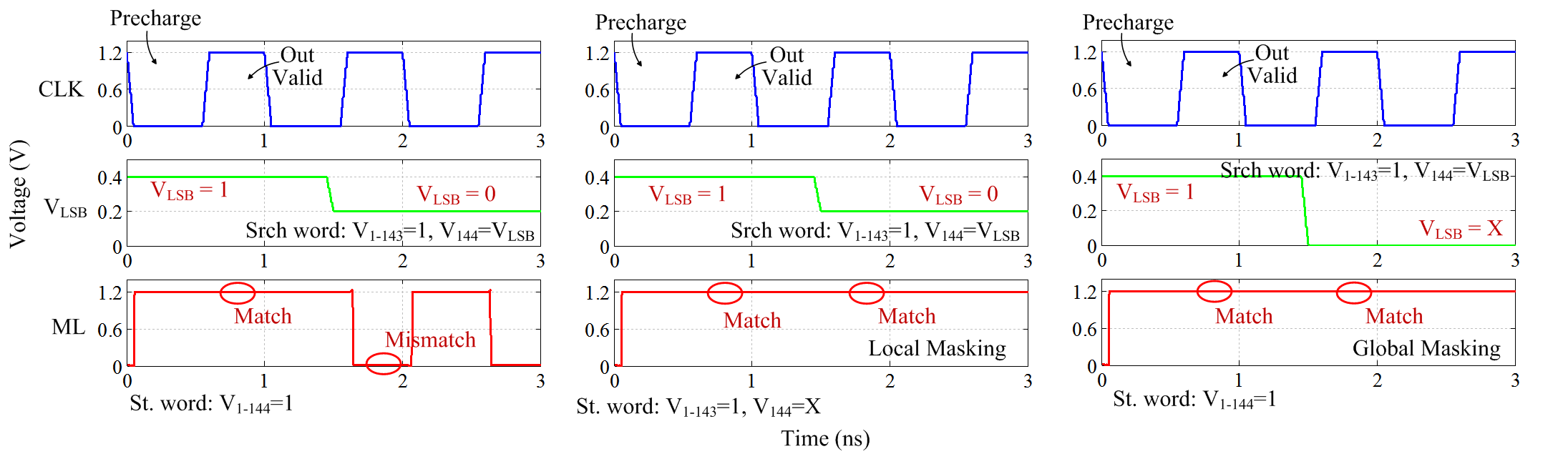}
         \caption{Transient waveforms for a 144-bit word search operation: (a, left) Match and mismatch. (b, middle) Local masking. (c, right) 
		 Global masking.}
          \label{}
        \end{figure*}

Standard CMOS-based static and dynamic TCAM cells are shown in Fig. 6(a-b) for comparison against the s-MTJ-based TCAM cells. The unconventional 
characteristic of the s-MTJ greatly reduces the footprint. Note that a standard CMOS-based static TCAM cell requires 16 
transistors and a dynamic TCAM cell requires six transistors and two trench capacitors. In contrast,
the skewed s-MTJ based design requires a {\it single} s-MTJ. Nonvolatile storage in the standard MTJs eliminates standby power dissipation
in the TCAM array. A narrow valley in the transfer characteristic of an s-MTJ (see Fig. 4) scales down the required minimum voltage difference 
between the stored logic `0' and `1', leading to a lower bias current through M1 and lower dynamic power in charging search lines 
($V_{2}$ capacitance). Moreover, only a single ended search word mapping (unlike CMOS designs) is needed, resulting in lower dynamic power 
dissipation and lower wiring overheads.

 TCAM cells in a column are connected in parallel to form the n-bit search/stored word. A TCAM array constitutes multiple such search columns 
 each storing a word from the database [Fig. 7(a)]. Fig. 7(b) shows the column sense-amplifier schematic comparing the resistance of 
 a s-MTJ-based 
 TCAM column against a reference resistance. Nodes OUT and OUTB in the sense-amplifier are charged to high in the precharge mode (CLK = 0). 
 Transistors M7-M10 equalize the node potential at their source and drain ends when CLK = 0. In the evaluate mode (CLK = 1), if the column 
 resistance due to TCAM cells is lower than the reference resistance ($R_{ref}$), the current through branch M5-M8 is higher leading to 
 OUT = 1 (Match). Otherwise, OUT = 0 (Mismatch), if the column resistance is higher.

 The reference resistance is configured such that it has a resistance in between the `worst case match' and the `worst case one-bit mismatch'. 
 For an n-bit search word, the `worst case match' column resistance $R_{match,n,w}$ obeys the relation
 \begin{equation}
 \frac{1}{R_{match,n,w}} = \frac{n}{R_{match,w}}
  \end{equation}
 where $R_{match,w}$ is the worst case match resistance in an s-MTJ-based TCAM cell. Note that for a match in the TCAM cell, the match 
 resistance ($R_{match}$) in s-MTJ follows $R_{match} \leq R_{match,w}$ for any combination of ternary bits: 0, 1, and X. Meanwhile, 
 the `worst case one-bit mismatch' column resistance $R_{mismatch,n,w}$ obeys the relation
  \begin{equation}
  \frac{1}{R_{mismatch,n,w}} = \frac{n-1}{R_{match,b}} + \frac{1}{R_{mismatch}}
   \end{equation}
where $R_{match,b}$ is the best case match resistance in an s-MTJ-based TCAM cell, and $R_{mismatch}$ is the mismatch resistance. 
Note that in case of a mismatch, the mismatch resistance $R_{mismatch}$ is always $\geq R_{missmatch,w}$ for any combination of ternary 
bits: 0, 1, and X. Figure 8 plots differential resistance between the `worst case match' and the `worst case one-bit mismatch', 
i.e., $\Delta R = R_{mismatch,n,w} - R_{match,n,w}$ at varying search word size ($n$). Note that $\Delta R < 0$ for $n > 20$. Thus, 
the s-MTJ characteristics limit the maximum number of TCAM cells in a column. The maximum number of parallel cells in a column can be 
increased by enhancing peak to valley resistance in s-MTJ (i.e., $R_{mismatch}/R_{match}$) and/or enhancing sharpness of the valley 
(i.e., by minimizing $R_{match,w} - R_{match,b}$). Nonetheless, large size search words can still be processed using s-MTJ-based TCAM cells 
by combining multiple block through an AND-tree as shown in Fig. 9(b), albeit at the cost of increasing peripheral area and power.

Operational waveforms for the TCAM array are shown in Fig. 10. At CLK = 0, the search bits $B_0$ and $B_1$ are decoded to search 
bit potential (0, 0.2, or 0.4 V depending on the search bit being `X', `0', or `1', respectively) using circuitry shown in Fig. 9(a). 
At CLK = 1, column peripherals in each block determine a match or mismatch and the following AND-tree combines their outputs to determine 
an n-bit (full length) match. Fig. 10 shows simulated transient of the TCAM array for a 144-bit search operation. As indicated in Fig. 10(a), 
the output (ML) is high in the case of a match between the search word and stored word, and ML becomes low with mismatch. Fig. 10(b) shows the 
case of a don’t care (`X') in the least significant bit (LSB) of the stored word. Therefore, in the search operation ML is high regardless of 
the LSB in search word (local masking). Similarly, when the search bit is `X', Fig. 10(c) shows a match irrespective of the corresponding 
stored bit (global masking). 

       \begin{figure}[!ht]%
       \centering
       \includegraphics[width=3in]{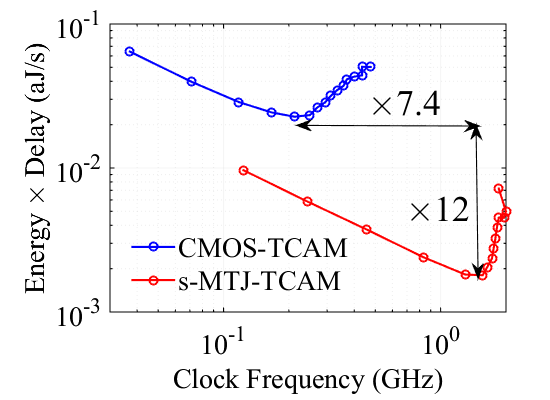}
       \caption{Comparison of Energy Delay Product(EDP) vs. clock frequency between s-MTJ-based TCAM and CMOS-based TCAM (TCAM array size 
	   144$\times$256) .}
       \label{}
       \end{figure}

In Fig. 11, at varying operational frequencies, the energy-delay-product (EDP) of the proposed s-MTJ-based dynamic TCAM is compared against 
that of CMOS-based dynamic TCAM. The energy-efficiency of s-MTJ-based TCAM is remarkably higher than that of CMOS-TCAM. Owing to a much more 
simplified cell and search operation, the minimum energy-delay product (EDP) is $\sim$12$\times$ smaller than the minimum EDP in CMOS-TCAM. 
Furthermore, while the operational frequency of CMOS-based TCAM is limited, the s-MTJ-based TCAM delivers a significantly improved performance. 
Note that the minimum EDP in s-MTJ-based TCAM occurs at $\sim$7$\times$ higher frequency than the minimum EDP frequency in CMOS-TCAM. At an 
operational frequency of $\sim$1 GHz, the EDP in sMTJ-based TCAM is $\sim$100$\times$ smaller than that in CMOS-TCAM. As interest in 
data-intensive and search-driven platforms (such as BigData) grows, the unique characteristics of an s-MTJ based TCAM will become 
increasingly important to reduce energ-delay product in such systems. 

\section{CONCLUSIONS}

This work has shown that the unique characteristics of a skewed s-MTJ can significantly simplify TCAM design and operation. In a skewed s-MTJ, 
the MTJ resistance can be controlled by the gate voltages $V_2$ and $V_3$. The resistance of an s-MTJ becomes maximum when $V_2$ and $V_3$ 
`match', i.e., they differ only by a fixed amount which we have called the `offset'. 
This associative property of the skewed s-MTJ enabled us to design a single s-MTJ-based match operation in a TCAM cell. The s-MTJ-based cell is 
minimized to one access transistor and one s-MTJ for a dynamic TCAM with global refresh. The cell is minimized to two transistors, two MTJs and 
one s-MTJ for dynamic TCAM with local refresh. The dynamic TCAM with local refresh has higher performance at the cost of slightly higher cell 
area. In the explored TCAM cell, the operation is non-Boolean and single ended which also minimizes dynamic power and routing. The s-MTJ-based 
cell shows $\sim$12$\times$ lower minimum energy-delay product (EDP) than CMOS-based cell. Moreover, the frequency at minimum EDP in the 
discussed cell is $\sim$7$\times$ higher than the frequency of minimum EDP in CMOS-based TCAM.  

\section*{Acknowledgement}

The work at Virginia Commonwealth University was supported by the US National Science Foundation under grant ECCS-1124714 and by the 
State of Virginia through the Research Commercialization Fund administered by the Center for Innovative Technology.

\end{document}